\documentclass{elsart}

\begin{document}

\begin{frontmatter}
\title{Probing additional dimensions in the universe
with neutron experiments}
\author[unam1,unam2]{Alejandro~Frank},
\author[ganil]{Piet~Van Isacker}
and 
\author[sevilla]{Joaqu\'\i n G\'omez-Camacho}

\address[unam1]{Instituto de Ciencias Nucleares,
Universidad Nacional Aut\'onoma de M\'exico,\\
Apartado Postal 70-543, 04510 M\'exico, D.F., M\'exico}
\address[unam2]{Centro de Ciencias F\'\i sicas,
Universidad Nacional Aut\'onoma de M\'exico,\\
Apartado Postal 139-B, 62251 Cuernavaca, Morelos, M\'exico}
\address[ganil]{GANIL, B.P.~55027, F-14076 Caen Cedex 5, France}
\address[sevilla]{Departamento de F\'\i sica At\'omica, Nuclear y Molecular,\\
Facultad de F\'\i sica, Universidad de Sevilla, Sevilla, Spain}

\begin{abstract}
We carry out a simple analysis of $(n+3)$-dimensional gravity
in the context of recent work on `large' supplementary dimensions
and deduce a formula for the expected compactification radius
for the $n$ additional dimensions in the universe,
as a function of the Planck and the electro-weak scales.
We argue that the correspondingly modified gravitational force
gives rise to effects that might be within the detection range
of dedicated neutron experiments.  
A scattering analysis
of the corresponding modified gravitational forces
suggests that slow neutron scattering off atomic nuclei
with null spin may provide an experimental test for these ideas.
\end{abstract}
\end{frontmatter}

The study of gravity at short range
has recently been the subject
of numerous theoretical and experimental investigations,
sparked by the proposal
by Arkani-Hamed, Dimoupoulos, and Dvali (ADD)~\cite{ADD98}
that gravity may depart from Newton's inverse square law at scales
which could be as large as a millimeter,
a scenario that was subsequently shown
to be consistent with string theory~\cite{ANT98}. 
Diverse experimental groups
have built refined versions of torsion balance experiments
and other ingenious designs to test gravity
at submillimeter ranges~\cite{HOY01}.
On the theoretical front, hundreds of papers have been written,
ranging from alternative formulations
of the large extra-dimensions framework (LED)~\cite{RAN99},
to the study of astrophysical constraints
and the expected experimental consequences
in future high-energy collider experiments
to black-hole production
and the effect of LED on fundamental symmetries.
But perhaps the most remarkable result to date
is the fact that no known physical constraints
have as yet falsified the LED theories.
>From the experimental point of view
the new measurements have confirmed
the validity of Newton's law to about 0.1~mm.
Although such efforts are of great value,
it can hardly be expected that this kind of experiments
can lower this limit significantly.

We show in this letter that
a simple analysis of $N$-dimensional ($N$-D) gravity
leads to compactification lengths
in close agreement with the more sophisticated calculations
of general relativity and string theory~\cite{BLU00}.
Following this cue, we then present calculations that
suggest that the effects of $N$-D forces,
although very small, might for particular values of $N$
fall within the detection range of dedicated neutron experiments.  

In their papers, ADD conjecture
the existence of two or more additional dimensions
in which gravity, but not the strong or the electro-weak forces,
might be acting,
diluting itself by spreading its lines of force
into these extra dimensions.
Essentially, this would explain its apparent weakness~\cite{ADD98}.
The proposal arises from a bold modification
of pre-existing many-dimensional string theories
and the more recent M-theories (M for membrane)
which encompass the former~\cite{HOR96}.
In these theories only gravitons
are able to traverse the extra dimensions,
whereas other particles are fixed to our observable 3-D world,
since the former are described as closed strings free to wander
while the latter are open strings with their ends fixed to our `brane'.
Additional dimensions are `compactified',
i.e., they are closed on themselves
with a characteristic radius of compactification $R_{\rm c}$
(which for simplicity is assumed
to be the same for all additional dimensions).
For ranges smaller than $R_{\rm c}$,
we thus expect a modified gravitational force.
The basic question is then:
Why should $R_{\rm c}$ be large
compared with M-theory's original Planck scale of $\sim10^{-35}$~m?
The answer rests on the empirical fact
that gravitation has never been measured below about $10^{-4}$~m
and more significantly, on the profound theoretical implications
that `large' extra dimensions would have
on the solution of the hierarchy problem
whose origin can be traced to the huge difference in strength
observed between gravity on the one hand
and the other forces described by the standard model~\cite{KAK93}.
In this scenario,
instead of catching up with the other forces at Planck's length scale,
the $N$-D gravitational force
actually joins the other interactions
at a distance about $10^{16}$ times larger,
namely at the electro-weak unification scale of $\sim10^{-19}$~m.
As will be shown below, this conjecture fixes
the strength of the $N$-D gravitational force
and the value of $R_{\rm c}$.
The most important consequence of the LED hypothesis
is the possible transition
from the entirely Platonic, inaccessible situation
posed by the Planck-scale compactification,
to one where it is conceivable
that experimental measurements may actually test these ideas.
The experiments envisioned to date are predominantly of two kinds.
The ones involving submillimeter torsion-balance experiments~\cite{HOY01}
mentioned above,
and high-energy collider experiments in the TeV energy region,
where diverse theoretical predictions exist
for the indirect observation of additional dimensions,
such as the occurrence of missing energy
carried away by undetected gravitons~\cite{CUL99}.
The question arises as to whether other experiments
can be designed to probe gravity at very short ranges.
It has been suggested, for example,
that the increasing precision
of experimental tests of the Casimir effect
may be used to probe new forces at micrometer distances~\cite{KRA02}.
Here we explore the possible effects of $N$-D gravity
in experiments with neutrons.

We start in the spirit of ADD~\cite{ADD98}
by carrying out a classical analysis of gravity
in $N=n+3$ dimensions.
If space would have $n+3$ (extended) dimensions,
Gauss' law implies that the force of gravity
would be of the form
\begin{equation}
F_n=-{{m_1m_2G_n}\over{r^{n+2}}},
\label{force}
\end{equation}
where $G_n$ is a constant
which reduces to Newton's gravitational constant $G$ for $n=0$.
As explained above, we shall follow M-theory
and assume that the masses,
as well as all other forces remain in 3-D space
and only the gravitational force field
leaks into the additional dimensions.
Even after compactification of these additional dimensions,
formula~(\ref{force}) should be correct for $r\leq R_{\rm c}$.

In general, one should assume a soft transition
from $N$-D gravity to 3-D gravity.
We shall instead follow a simpler procedure
and consider a sudden transition from $N$-D to 3-D
using Eq.~(\ref{force}).
We first impose the equality of forces
at the compactification length $R_{\rm c}$,
$F_n(R_{\rm c})=F_0(R_{\rm c})$,
which implies that
\begin{equation}
F_n=-{{m_1m_2GR_{\rm c}^n}\over{r^{n+2}}}.
\label{force2}
\end{equation}
We then implement the ADD conjecture
that the $N$-D force at the electro-weak length $R_{\rm e}$
is as strong as the $3$-D force at the Planck length $R_{\rm P}$,
$F_n(R_{\rm e})=F_0(R_{\rm P})$.
This leads to the desired equation
\begin{equation}
\left({{R_{\rm c}}\over{R_{\rm e}}}\right)^n=
\left({{R_{\rm e}}\over{R_{\rm P}}}\right)^2,
\label{conj}
\end{equation}
which expresses the compactification length $R_{\rm c}$
as a function of the two fundamental scales $R_{\rm e}$ and $R_{\rm P}$.
These lengths are defined as
$R_{\rm P}=\sqrt{{\hbar G}/{c^3}}$
and $R_{\rm e}=\sqrt{G_{\rm F}/\hbar c}$,
where $G$ and $G_{\rm F}$ are Newton's and Fermi's constants, respectively,
the latter defined by~\cite{KAK93}
$G_{\rm F}={{\sqrt{2}\hbar^2g^2}/{8c^2M_W^2}}=
1.166\cdot10^{-5}GeV^{-2}(\hbar c)^3$,
where $M_W$ is the mass of the $W_\pm$ bosons
and $g\sin\theta_W=e$,
with $\theta_W$ the Weinberg angle.
We use cgs units throughout.
We then find that formula~(\ref{conj}) can be written in the form:
\begin{equation}
\left({{R_{\rm c}}\over{R_{\rm e}}}\right)^n=
{{c^2}\over{\hbar^2}} {{G_{\rm F}}\over G}.
\label{conj2}
\end{equation}
Note that this expression relates
the number of additional dimensions $n$,
the radius of compactification $R_{\rm c}$,
and the electro-weak scale $R_{\rm e}$,
to the ratio of two fundamental numbers in nature:
Fermi's and Newton's constants,
with $c^2/\hbar^2$ as a proportionality constant.

To have an indication of whether formula~(\ref{conj2}) is robust,
we remark that a similar result can be obtained
using a different argument.
We require that at the electroweak length scale
the $N$-D force be comparable to the electromagnetic interaction.
We may assume that at distances of order $R_{\rm e}$
most particles are ultra-relativistic
and thus that their masses should be of order $\hbar/R_{\rm e}c$.
Equating the electromagnetic interaction to the gravitational force
between two masses of this order
we arrive at
\begin{equation}
{{\hbar^2}\over{R_{\rm e}^2 c^2}}
{{GR_{\rm c}^n}\over{R_e^{n+2}}}=
{{e^2}\over{R_{\rm e}^2}}.
\label{ndg=el}
\end{equation}
This leads to the relation
\begin{equation}
\left({{R_{\rm c}}\over{R_{\rm e}}}\right)^n=
\alpha{{c^2}\over{\hbar^2}} {{G_{\rm F}}\over G}=
\alpha\left({{R_{\rm e}}\over{R_{\rm P}}}\right)^2,
\label{conj3}
\end{equation}
where $\alpha$ is the fine-structure constant
which is about $1/128$ at the electroweak scale $R_{\rm e}$.
Values for $R_{\rm c}$
obtained with expressions~(\ref{conj2}) and (\ref{conj3})
are compared in Table~\ref{tab}.

The present model can be interpreted as follows:
The Fermi constant is related
in the standard electro-weak theory
to the value of the scalar field
that produces the Higgs mechanism of spontaneous symmetry breaking.
The Higgs boson, which is yet to be discovered,
corresponds to the excitation mode of this scalar field.
However, in this model $G_{\rm F}$ also represents
the intensity of the gravitational field in $N$ dimensions.
Thus, this model suggests the exciting possibility
that the scalar field which is needed in the standard model
may just be the gravitational field
which is wrapped out into the additional dimensions~\cite{GIU01}.
The very small Newton constant $G$
gives the residual value of the gravitational field
which spills out into the usual three dimensions,
beyond the radius of compactification $R_{\rm c}$.

Carrying further this line of thinking,
we can consider that, for distances larger than $R_{\rm c}$,
the gauge forces live in the 3-D dimensional brane,
and so does the normal, very weak gravity.
Between $R_{\rm c}$ and $R_{\rm e}$,
the gauge forces still live in three dimensions,
while gravity lives in $N$ dimensions,
increasing its strength as the distance decreases,
so that for $r=R_{\rm e}$ is becomes comparable
to the electro-weak force.
At this point, and due to a yet unknown mechanism,
gravity generates the scalar field
that couples to the gauge field
giving rise to the Higgs mechanism.

In Table~\ref{tab} we display both values of $R_{\rm c}$
as a function of $n$,
which turn out to be close to the ones
evaluated by other means.
\begin{table*}
\caption{\label{tab}
Estimates of various lengths (in fm),
energies (in MeV),
and phase shifts (in rad)
as a function of the number of extra dimensions $n$.}
\begin{tabular}{lccccccc}
&&&&&&&\\
\hline
$n$&1&2&3&4&5&6&7\\
\hline
$R_{\rm c}$ (Eq.~\ref{conj})&
$1.2~10^{30}$&
$2.8~10^{13}$&
$8.1~10^7$&
$1.4~10^5$&
$3.0~10^3$&
$2.3~10^2$&
$38$\\
$R_{\rm c}$ (Eq.~\ref{conj3})&
$9.3~10^{27}$&
$2.5~10^{12}$&
$1.6~10^7$&
$4.2~10^4$&
$1.1~10^3$&
$1.0~10^2$&
$19$\\
$V_n(r=30)$&
$1.6~10^{-7}$&
$2.4~10^{-12}$&
$4.0~10^{-17}$&
$7.1~10^{-22}$&
$1.3~10^{-26}$&
$2.6~10^{-31}$&
$5.1~10^{-36}$\\
$R_=$&
$18$&
$25$&
$32$&
$39$&
$47$&
$54$&
$62$\\
$V_n(R_=)$&
$4.4~10^{-7}$&
$4.1~10^{-12}$&
$3.1~10^{-17}$&
$1.9~10^{-22}$&
$9.0~10^{-28}$&
$4.2~10^{-33}$&
$1.5~10^{-38}$\\
$E_{\rm min}$&
---&
---&
$2.3~10^{-13}$&
$7.9~10^{-8}$&
$1.7~10^{-4}$&
$2.7~10^{-2}$&
$1.1$\\
$E_{\rm opt}$&
$4.6$&
$2.3$&
$1.5$&
$0.98$&
$0.68$&
$0.51$&
$0.39$\\
$\phi_{\rm opt}$&
$7.1~10^{-8}$&
$1.2~10^{-12}$&
$1.6~10^{-17}$&
$2.1~10^{-22}$&
$4.9~10^{-28}$&
$5.6~10^{-33}$&
$1.2~10^{-38}$\\
\hline
\end{tabular}
\end{table*}
Note that $n=1$ can be readily discarded
since it leads to a value of $R_{\rm c}$
larger than the size of the solar system
and hence to unstable planetary orbits.
For $n=2$ we find $R_{\rm c}\sim$ mm or cm.
Deviations from Newton's law at this scale
seem to be discarded by experiment.
Nevertheles, it should be noted that our analysis
can only be expected to give rough estimates
with considerable uncertainties
in the prediction of $R_{\rm c}$.
It is the range below 1~mm,
which is very difficult to explore with macroscopic gravity experiments,
that we would like to investigate by means of neutrons.
We should note that even if the LED hypothesis turns out to be wrong,
it is still an interesting question to analyze
whether neutron experiments can unveil deviations
from Newton's law at short distances.

The physics of slow neutrons
has undergone significant evolution in the last decades.
Neutrons have become a standard probe
for nuclear physics experiments
as well as for other areas
including the study and dynamics of condensed matter~\cite{BYR94}.
Pulsed neutron beams can currently be generated
with very precise energies and polarizations
and neutron detectors achieve very high efficiencies.
Delicate experiments with thermal neutrons
have recently demonstrated the quantization of their energies
when subject to earth's gravitational field~\cite{NES02}.
In order to attempt neutron $N$-D gravity experiments
at short ranges we face two problems from the outset.
The more obvious one is the strong nuclear force,
present at range scales of the order of $10^{-15}$~m.
A second, less obvious problem is
that even for a spin-zero target nucleus,
a neutron approaching it with speed $v$
feels a magnetic field $\vec B={1\over{2c}}\vec E\times\vec v$
in its rest frame,
where $\vec E$ is the nuclear electric field
due to its charge $Z$.
A long-range electromagnetic interaction
(the Schwinger effect~\cite{ALE01})
of strength $\vec\mu\cdot\vec B$ results,
where $\vec\mu$ is the neutron magnetic moment.
Neutrons have been proposed primarily to avoid direct competition
with the much stronger electromagnetic interaction.
We see that there is a residual, relativistic effect
which needs to be dealt with.
To minimize this potentially competing interaction
slow neutrons are required,
possibly polarized in the incident direction.
(Note that the effect averages out to zero for unpolarized projectiles.)
Very slow neutrons will suffer essentially
pure $s$-wave nuclear scattering,
while the longer range $N$-D gravitational force
can in principle produce scattering of higher $l$-waves.
The main question is whether interference effects
between nuclear and gravitational scattering can be detected,
in a fashion similar to the observed interference effects
between nuclear and electromagnetic forces in heavy-ion reactions
which give rise to `rainbow' scattering
and other such phenomena~\cite{VIL93}.

The potential that produces the modified gravitational force
for distances $r$ below $R_{\rm c}$ can be written as 
\begin{equation}
V_n(r)={{m_1m_2GR_{\rm c}^n}\over{(n+1)r^{n+1}}}.
\label{potential}
\end{equation}
Let us first consider the interaction of a neutron beam
with a heavy nucleus such as $^{208}$Pb.
The constant $m_1m_2G$ is extremely small in this case
giving rise to very small values of the potential energy
at a typical distance of $r=30$ fm (see Table~\ref{tab}).
We will discuss what are the optimal experimental conditions
which could allow observation of  this tiny effect
in neutron-scattering experiments
and will consider what is the adequate energy and angular momentum
so that the phase shift due to the gravitational force
is as large as possible.

The nuclear potential can be parametrized
with a Woods-Saxon shape, so that
\begin{equation}
V_{\rm nucl}(r)={{V_0}\over{1+\exp(r-R)/a}},
\label{ws}
\end{equation}
Reasonable parameters are
$V_0=50$ MeV,
$R=1.2~A^{1/3}$ fm,
and $a=0.6$ fm.
This gives, for distances below $r=10$ fm,
values of the potential in the MeV range
and any gravitational effect at that distance
would be drowned by the uncertainties in the nuclear potential.
Instead, one must probe distances
at which the nuclear and gravitational potential
are of the same order.
In Table~\ref{tab} we indicate  the distances $R_=$
at which the nuclear and gravitational potential are equal,
as a function of $n$.
For these calculations we have used the estimates for $R_{\rm c}$
from Eq.~(\ref{conj2})
although Eq.~(\ref{conj3}) leads to similar results.
Any scattering observable
that is affected by distances smaller than $R_=$
will be contaminated by nuclear effects.
We find that,
if the number of additional dimensions is larger than $n=6$,
then the gravitational force will be smaller
than the nuclear force for any value of $r<R_{\rm c}$.
For $n\leq6$ there exists a range $R_=<r<R_{\rm c}$
in which the gravitational effects
are larger than nuclear effects,
and, at least in principle, might be measurable although small.

The scattering observables are also affected by the fact that
the neutron has bound states in the nuclear potential
generated by $^{208}$Pb.
>From the shell structure of this nucleus
one knows that the single-particle potential
supports bound states up to angular momentum $L=7$
(the $1j_{15/2}$ orbital).
Thus, the scattering of neutrons with $L\leq7$
is affected by the nuclear potential,
even if the scattering is very small,
because the scattering wave functions
have to be orthogonal to the bound states.
Consequently, to obtain scattering observables
free of nuclear contamination,
we need to consider $L>7$.

In order to investigate gravitational effects,
the energy of the neutron cannot be arbitrarily low.
As one is investigating effects
which occur at distances below the compactification length $R_{\rm c}$,
the wave length of the neutron should be smaller than $R_{\rm c}$.
More specifically,
the momentum of the neutron should be such that
\begin{equation}
p_{\rm n}R_{\rm c}>(L_{\rm min}+1/2)\hbar,
\end{equation}
where $L_{\rm min}=8$ in the example of $^{208}$Pb.
The corresponding minimum energies of the neutron,
\begin{equation}
E_{\rm min}=
{{p^2_{\rm n}}\over{2m_{\rm n}}}=
{{(L_{\rm min}+1/2)^2\hbar^2}\over{2m_{\rm n}R^2_{\rm c}}},
\end{equation}
are given in Table~\ref{tab}.

Estimates of the phase shifts
due to the gravitational interaction
can be found in the eikonal approximation
where they are obtained for a given angular momentum
in terms of the time integral of the potential
along a straight line trajectory
which has the same angular momentum:
\begin{equation}
\phi(L,E)\simeq
{1\over\hbar}\int V(\sqrt{b^2+v^2t^2})dt,
\qquad
b={{\hbar(L+1/2)}\over{m_{\rm n}v}}.
\end{equation}
The time integral can be estimated
by taking the potential at the point of closest approach $r=b$, 
multiplied by the approximate characteristic time $\tau=b/v$ of interaction.
In this way, we find
\begin{equation}
\phi(L,E,n)\simeq
{{m_1m_2GR_{\rm c}^n}\over{(n+1)b^n\hbar v}}.
\label{phi}
\end{equation}
>From this expression we see that,
in order to enhance the scattering effects
of the gravitational force,
one would need to have, in principle,
small impact parameter and small velocity
which are related through $bvm_{\rm n}=(L+1/2)\hbar$.
To maximize the phase shift $\phi(L,E,n)$,
the best choice is to take
the minimum angular momentum $L_{\rm min}=8$
and the minimum impact parameter
which corresponds to the distance $R_=$
at which the nuclear force may start to play a role.
This gives the following optimal energy for scattering:
\begin{equation}
E_{\rm opt}=
{{(L_{\rm min}+1/2)^2\hbar^2}\over{2m_{\rm n}R^2_=}}.
\end{equation}
These energies are shown in Table~\ref{tab}
and are of the order of 1 MeV.
At these energies the optimal phase shifts (for $L=L_{\rm min}$),
also shown in Table~\ref{tab},
can be evaluated to give
\begin{equation}
\phi_{\rm opt}\simeq
{{V_n(R_=)}\over{2E_{\rm opt}}}.
\end{equation}
For $n>2$ the phase shift is extremely small and decreases as $n$
grows. However, we believe that the case of $n=2$ may be within
reach of current dedicated scattering experiments. For this case, the
elastic scattering amplitude produced by the modified
gravitational force is given by the expression
\begin{equation}
A_g(E,\theta) = {i \over 2 k} \sum_L \phi(L,E,2) P_L(\cos \theta)
\end{equation}
where $\phi(L,E,2)$ is given by Eq.~\ref{phi}. This amplitude
turns out to be energy independent and can be written in the
closed form
\begin{eqnarray}
A_g(E,\theta) &=& i A_g f_g(\theta),
\nonumber\\
A_g           &=& {2 \hbar c (m_n m_T G_F)^2 \over 3 (m_n+ m_T)},
\nonumber\\
f_g(\theta) &=& \sum_L (L+1/2)^{-1} P_L(\cos \theta),
\label{ampli}
\end{eqnarray}
where we have made use of Eq.~\ref{conj2}. The modified
gravitational amplitude  can now be evaluated for the scattering
of neutrons on $^{208}$Pb, using Eq.~(\ref{ampli}) and we find $A_g=
0.298 \cdot 10^{-8}$ fm, to be compared with the typical
scattering amplitudes for the nuclear force, which for a range of
energies of a few MeV, are of the order of $A_n = 7$ fm, although
the amplitude can strongly fluctuate with energy as resonances are
crossed. It would seem that it is impossible to observe such a
tiny gravitational effect, being so small compared to the nuclear
amplitude. However, the angular dependence of these amplitudes is
quite different. In contrast to the nuclear part, which is
essentially independent of the scattering angle for $\theta \ll
1/L_{min}$, the gravitational amplitude involves the contribution
of a significant number of angular momenta. This in turn implies
that the gravitational amplitude diverges for small scattering
angles. It is the combination of these characteristics which may
open a window to observe an interference effect. More
specifically, we have derived that $f_g(\theta) = K_0(\theta/2) +
\delta_g(\theta)$, where $K_0$ is the Bessel function, which
diverges logarithmically as $\theta\to 0$, and $\delta_g(\theta)$
is a smooth function of the angle.

Our analysis shows that at very small scattering angles,
the neutron-nucleus differential cross section is given by
\begin{eqnarray}
{d \sigma \over d \Omega} &=& |A_n(E,\theta) + A_g(E,\theta)|^2 \\
  &\simeq& |A_n(E,0)|^2 + 2 |A_n(E,0)| A_g \sin(\arg(A_n(E,0)) K_0(\theta/2)
\end{eqnarray}
Note that while $|A_n(E,0)|^2$ is about 5 b/sr, $2 |A_n(E,0)| A_g$
is about 0.4 nb/sr. Discerning such a faint whisper in the midst
of the nuclear background roar can be a formidable task. But this
feat may be accomplished by carefully monitoring both the angular
and energy dependence of the cross section. As the phase of the
nuclear amplitude changes as resonances are crossed, the
diminutive interference between nuclear and gravitational
amplitudes changes from constructive to destructive interference.
Our proposal thus requires measurements involving two
detectors, one at the smallest possible angles and a second at
slightly larger ones. For purely nuclear scattering the ratio of
the cross sections (the ratio of detected neutrons) should remain
constant as energy is changed. The presence of a gravitational
effect of the kind discussed here would be signaled by small
fluctuations associated to the interference between the two
interactions. Moreover, these fluctuations would not be random,
but should correlate with the magnitude of the cross section. A
careful analysis of these fluctuations may isolate a gravitational
signature.

Considering the progress that has been achieved recently in
neutron physics, as shown, for example, in time-of-flight
experiments measuring the precession of polarized neutrons through
a gas~\cite{FRA01}, these ideas could be tested in the near
future.

Although we have concentrated here
on the effects of extra-dimensional gravity
on neutron scattering, an entirely different approach
can be attempted~\cite{NES}.
The quantum  effects of earth's gravity  on neutrons
have been observed in experiments
by Nesvizhevsky {\it et al.}~\cite{NES02}.
The interference pattern of the neutron density,
which is of $\mu$m scale,
could be affected by deviations from Newton's law at $\mu$m scale,
produced by the modified  gravity of the plates.
For $n=2$ 
the estimate of $R_{\rm c}$ is in the cm range, so
$N$-D gravity, at the $\mu$m range, is 
$10^8$ times larger than normal gravity.
The current experimental setup~\cite{NES02}
could discern effects for forces $10^{10}$ times larger
than normal gravity~\cite{NES}. 
Both kinds of dedicated neutron experiments
could shed light on the quest for additional dimensions in the universe. 

We wish to thank H.G.~B\"orner, R.F.~Casten, V.~Nesvizhevsky
and A.~Villari for encouraging discussions.
AF is supported by CONACyT, Mexico, 
and JGC by the spanish DGICyT project FPA2002-04181-C04-04.

\end{document}